\documentclass[11pt]{article}
\usepackage[english,activeacute]{babel}
\usepackage{epsfig}
\usepackage{amsmath}
\usepackage{epic}
\usepackage{amssymb}
\usepackage{fancybox}
\usepackage{wrapfig}
\usepackage[font=footnotesize,labelsep=period]{caption}
\usepackage{amsthm}
\usepackage{float}
 \usepackage[linktocpage=true,colorlinks]{hyperref}
 \usepackage[capitalise]{cleveref}
\usepackage{xcolor}
\definecolor{dark-red}{rgb}{0.6,0.15,0.15}
\definecolor{dark-blue}{rgb}{0.15,0.15,0.8}
\definecolor{medium-blue}{rgb}{0,0,0.6}
\hypersetup{
    colorlinks, linkcolor={dark-red},
    citecolor={dark-blue}, urlcolor={medium-blue}
}
\allowdisplaybreaks
\usepackage{booktabs}

\usepackage{wrapfig}
\usepackage{cleveref}

\newcommand{\dashedrightarrow}[2][]{\ext@arrow 0359\rightarrowfill@@{#1}{#2}}

\usepackage[all]{xy}
\usepackage{xypic, color, graphics}

\theoremstyle{plain}

\theoremstyle{definition}

 \def\v #1{\vert #1\vert}             
 
 \def\m #1 #2{(-1)^{{\v #1} {\v #2}}} 

 \let \m=\medskip

\def\bea{\begin{eqnarray}}
 \def\eea{\end{eqnarray}}
\newcommand{\beq}{\begin{eqnarray}}
\newcommand{\eeq}{\end{eqnarray}}
\newcommand{\ba}{\begin{array}}
\newcommand{\ea}{\end{array}}
\newcommand{\be}{\begin {equation}}
\newcommand{\ee}{\end{equation}}
\def\picture #1 by #2 (#3){
  \vbox to #2{
    \hrule width #1 height 0pt depth 0pt
    \vfill
    \special{picture #3} 
    }
  }
\def\scaledpicture #1 by #2 (#3 scaled #4){{
  \dimen0=#1 \dimen1=#2
  \divide\dimen0 by 1000 \multiply\dimen0 by #4
  \divide\dimen1 by 1000 \multiply\dimen1 by #4
  \picture \dimen0 by \dimen1 (#3 scaled #4)}
  }
  
\usepackage{fancyhdr}

\usepackage[Bjornstrup]{fncychap}
\usepackage{fancyhdr}

\usepackage{appendix}
\usepackage[nottoc,numbib]{tocbibind}

\begin{document}

\centerline{\Large \bf Integrable $1+1$ dimensional hierarchies arising }\vskip 0.25cm
\centerline{\Large \bf from reduction of a non-isospectral problem in $2+1$ dimensions}
\vskip 0.5cm
\centerline{P.G. Est\'evez\dag,\ J.D. Lejarreta\ddag, \ C. Sard\'on\dag}
\vskip 0.5cm
\centerline{\dag Department of Fundamental Physics, University of
Salamanca,}
\centerline{Plza. de la Merced s/n, 37.008, Salamanca, Spain.}
\centerline{\ddag Departamento de F\'{\i}sica Aplicada, Universidad de Salamanca,}
\centerline{Plza. de la Merced s/n, 37.008, Salamanca, Spain.}
\vskip 1cm
\begin{abstract}
This work presents a classical Lie point symmetry analysis of a two-component, non-isospectral Lax pair of a hierarchy of partial differential equations in $2+1$ dimensions, which can be considered as a modified version of the Camassa-Holm hierarchy in $2+1$ dimensions. A classification of reductions for this spectral problem is performed. Non-isospectral reductions in $1+1$ dimensions are considered of remarkable interest.
\end{abstract}

\section{Introduction}
Solving non-linear partial differential equations (PDEs) can be a difficult task. Several procedures
have been developed in order to solve them \cite{Ames}, \cite{Hill}, \cite{Olver}, \cite{Ovs}, \cite{Steph}. The integrability of non-linear PDEs can be guaranteed by means of the Painlev\`e test \cite{Conte}, \cite{EP}, \cite{WTC}, the existence of a Lax Pair \cite{EP2}, or the inverse scattering transform  \cite{AC}. Also, the use of geometric techniques has also proved to be very successful for the integration of differential equations \cite{CL}.
\bigskip

Lie symmetries were first introduced by Lie \cite{Steph} in order to solve ordinary differential equations (ODEs) or
reduce a system of equations to a simpler form \cite{BC}, \cite{Olver}.
Lie point symmetries represent a powerful tool, although they involve lengthy calculations for the most part. Nevertheless, the interest in them has grown in the past years due to the development of symbolic computation packages. There exist different types of symmetry transformations. Here, we shall focus on \textbf{Lie point symmetries}, where the coefficients of the infinitesimal generator just depend on the coordinates and fields. Each Lie point symmetry leads us to a  reduced version of the equation with the number of independent
variables diminished by one. This fact implies that any solution of a PDE can be derived by reduction of the same to an ODE. Lie point symmetries can be classified as classical \cite{Steph} and non-classical
\cite{BC}, \cite{Olver}. Both approaches constitute the usual way to identify the reductions. The case that concerns us will be based on the classical approach.
\bigskip

Much less frequent is the identification of  Lie symmetries of the spectral problem of a PDE system \cite{Leg} although the introduction of an spectral (or non-isospectral) parameter in a linear problem without spectral parameter through one-parametric groups of Lie points symmetries was proposed by Levi et al \cite{levi1}, \cite{levi2}. This group interpretation of the spectral parameter has been extensively studied by Cie\'sli\'nski in several papers \cite{cies1}, \cite{cies2}, \cite{cies3}, \cite{cies4}, One of us and collaborators have proposed such procedure before in \cite{EGP} and have continued along this line of research \cite{BEGP}, \cite{EGL}, \cite{ELS}. So have other authors contributed to it \cite{Zhi}. It is important to remark that the study of symmetries of a Lax pair provides a lot of more information than the simple identification of the symmetries of a set of PDEs, as it allows us not just to determine the reduction of the fields, but also that of the eigenfunctions and the spectral parameter. The non-isospectral case is quite relevant. Under this premise, the spectral parameter has to be considered as an additional field and sometimes, the reduced spectral problem is non-isospectral as well.

In the present paper, we are concerned with a non-isospectral problem. This non-isospectral problem was first introduced in
\cite{PGE4} as an integrable generalization of the Qiao hierarchy \cite{Qiao} to $2+1$ dimensions. This hierarchy has been proved to be connected (through reciprocal transformations) \cite{PGE4}, \cite{ES} with the Camassa-Holm hiertarchy in 2+1 dimensions \cite{EP3} and it  shall be
denoted under the name mCH(2+1) (modified Camassa-Holm hierarchy in 2+1 dimensions) henceforth in this article.

The plan of the paper is the following: Section 2 is devoted to the descripcion of mCH(2+1) and its non-isospectral Lax pair. In section $3$ we apply the classical Lie method for finding Lie point symmetries of the Lax pair of mCH(2+1).
 Section $4$ contains all the possible
reductions under the symmetries identified in section $3$. Eight non-trivial spectral problems in 1+1 dimensions and their correspondent reduced hierarchies arise from such reductions. Six of them are non-isospectral, the other two are isospectral. We shall close this paper with a summary of the most relevant results.
\section{The  mCH(2+1) Hierarchy}
\subsection{Lax pair}
In \cite{PGE4}, a generalization to $2+1$ dimensions of the Qiao hierarchy \cite{Qiao}, \cite{Qiao10} was
presented. Upon the assistance of reciprocal transformations, mCH(2+1) was proved to be
equivalent to $n$ copies of the modified Calogero-Bogoyavlenskii-Schiff equation (CBS) in three dimensions \cite{Bogo}, \cite{Calo}, \cite{EP2}, \cite{KP}.
Each copy depends on three different variables and has the Painlev\'e property
\cite{EP2}, \cite{KP}. Their correspondent Lax pair was proposed in \cite{EP2}. The spectral problem for mCH(2+1)
is obtained by means of the inverse reciprocal transformation \cite{PGE4}. This spectral problem corresponds with the
following two component
non-isospectral Lax pair which contains $2n+1$ fields. The spectral parameter will also be considered as a field:

\begin{equation}
\left(\begin{array}{cc}
 \phi\\
 \psi
\end{array}\right)_x=\frac{1}{2}\left[\begin{array}{cc}
 -1& i\sqrt{\lambda}\,u\\
 i\sqrt{\lambda}\,u& 1
\end{array}\right]\left(\begin{array}{cc}
 \phi\\
 \psi
\end{array}\right),\label{2.1}
\end{equation}
\begin{eqnarray}
\left(\begin{array}{c}
 \phi\\
 \psi
\end{array}\right)_t=\lambda^{n}\left(\begin{array}{c}
 \phi\\
 \psi
\end{array}\right)_y+\lambda \,p \left(\begin{array}{c}
 \phi\\
 \psi
\end{array}\right)_x+\frac{i\sqrt{\lambda}}{2}\left[\begin{array}{cc}
 0&q_{x}-q\\
 q_{x}+q& 0
\end{array}\right]_x\left(\begin{array}{c}
 \phi\\
 \psi
\end{array}\right).\label{2.2}
\end{eqnarray}
where:
$$u=u(x,y,t),\quad\quad \lambda=\lambda(y,t)$$
$$
p=p(x,y,t)=\sum_{j=1}^n{\lambda^{(n-j)}(y,t)\,\omega^{[j]}}(x,y,t),$$ $$ q=q(x,y,t)=\sum_{j=1}^n{\lambda^{(n-j)}(y,t)\,v^{[j]}}(x,y,t).
$$
and  $i=\sqrt{-1}$.

\subsection{Non-isospectrality and equations for the hierarchy}
The compatibility condition between (\ref{2.1}) and (\ref{2.2}) yields
the non-isospectral
 condition: \begin{equation}\lambda_t-\lambda^n\lambda_y=0\label{2.3}\end{equation} as well as the equations:
\begin{align}
\nonumber &\left(v_{xx}^{[n]}-v^{[n]}\right)_x-u_t=0,&\\
\nonumber &v_{xx}^{[j]}-v^{[j]}+u\omega^{[j+1]}=0,\qquad &j=1\dots n-1,\\
\nonumber &\left(u\omega^{[1]}\right)_x+u_y=0,&\\
&\omega_x^{[j]}=uv_x^{[j]}, \qquad &j=1\dots n.\label{2.4}\end{align}
\subsection{Recursion operator and hierarchy}
The system of equations (\ref{2.4}) can be expressed in a more compact form in terms of the operators $K$ and $J$, defined as:
\begin{equation}K=\delta^3-\delta,\qquad J=-\delta u\delta^{-1} u\delta,\qquad \delta=\frac{\partial}{\partial x}\label{2.5}.\end{equation}
$R$ shall be the recursion operator characterized as: \begin{equation}R=JK^{-1}\Rightarrow R^{-1}=KJ^{-1}.\end{equation}
Thus, the $n$ dimensional hierarchy can be written in terms of the following expression:
\begin{equation}u_t=R^{-n}u_y.\end{equation}
The $1^{st}$ and $n^{th}$ fields appear through:
\begin{equation}u_y=Jv^{[1]},\qquad u_t=Kv^{[n]}\end{equation}
and the recurrence relation for the $j^{th}$ fields is:
\begin{equation}Jv^{(j+1)}=Kv^{[j]} \Rightarrow v^{(j+1)}=J^{-1}Kv^{[j]}, \qquad j=1\dots n-1.\end{equation}

\section{Classical symmetries for the spectral problem of  mCH(2+1) }
\subsection{Lie Point Symmetries}
We are looking for Lie point symmetries of the spectral problem presented in section $2$.
Symmetries of system (\ref{2.4}) are interesting in themselves, but we find important to suggest how
\textbf{the eigenfunction and the spectral parameter transform under the action of a Lie symmetry}.
More precisely, we wish to know how these fields look when reduced by the symmetry to $1+1$ dimensions. For this reason, we shall apply the Lie method to the spectral problem (\ref{2.1})-(\ref{2.3}), instead of applying it to the system
 of equations in (\ref{2.4}).

We shall proceed by writing the infinitesimal Lie point transformation for the variables and fields.
 It is important to note that the spectral parameter $\lambda(y,t)$ is not a constant
(see (\ref{2.3})). Therefore, it has to be considered as an additional field, which implies that we not only have
to look for symmetries of the Lax pair, but also for the non-isospectral condition (\ref{2.3}).
We have proven the benefits of such a procedure \cite{Leg} in a previous
paper \cite{EP2}.

The uniparametrical Lie point transformation proposed is:
\begin{align}
\hat x&=x+\epsilon\,  \xi_1(x,y,t,u,\omega^{[i]},v^{[i]},\phi,\psi)+O(\epsilon^2),\nonumber \\
\hat y&=y+\epsilon\,  \xi_2(x,y,t,u,\omega^{[i]},v^{[i]},\phi,\psi)+O(\epsilon^2),\nonumber \\
\hat t&=t+\epsilon\,  \xi_3(x,y,t,u,\omega^{[i]},v^{[i]},\phi,\psi)+O(\epsilon^2),\nonumber \\
\hat u&=u+\epsilon\,  \eta_u(x,y,t,u,\omega^{[i]},v^{[i]},\phi,\psi)+O(\epsilon^2),\nonumber \\
\hat \omega^{[j]}&=\omega^{[j]}+\epsilon\,  \eta_{\omega}^{[j]}(x,y,t,u,\omega^{[i]},v^{[i]},\phi,\psi)+O(\epsilon^2),\nonumber \\
\hat v^{[j]}&=v^{[j]}+\epsilon\,  \eta_v^{[j]}(x,y,t,u,\omega^{[i]},v^{[i]},\phi,\psi)+O(\epsilon^2),\nonumber \\
\hat \lambda&=\lambda+\epsilon\,  \eta_{\lambda} (x,y,t,\lambda, u,\omega^{[i]},v^{[i]},\phi,\psi)+O(\epsilon^2),\nonumber \\
\hat \phi&=\phi+\epsilon\,  \eta_{\phi}(x,y,t,\lambda,u,\omega^{[i]},v^{[i]},\phi,\psi)+O(\epsilon^2),\nonumber \\
\hat \psi&=\psi+\epsilon\,  \eta_{\psi}(x,y,t,\lambda,u,\omega^{[i]},v^{[i]},\phi,\psi)+O(\epsilon^2)\label{3.1}\end{align}
where $i,j=1\dots n.$

The Lie algebra corresponds to the set of vector fields of the form:
\begin{align}
X=\xi_1\frac{\partial}{\partial x}+\xi_2\frac{\partial}{\partial y}+\xi_3\frac{\partial}{\partial t}+ \eta_{\lambda}\frac{\partial}{\partial \lambda}+\eta_u\frac{\partial}{\partial u}+\sum_{j=1}^{n}{\eta_{\omega}^{[j]}\frac{\partial}{\partial \omega^{[j]}}}+\nonumber\\
\quad+\sum_{j=1}^{n}{\eta_v^{[j]}\frac{\partial}{\partial v^{[j]}}}+\eta_{\phi} \frac{\partial}{\partial \phi}+\eta_{\psi} \frac{\partial}{\partial \psi}
.\label{3.2}
\end{align}

At this point, it is essential to use the concept of \textbf{prolongation of an action of a Lie group}, as our spectral problem contains derivatives of the fields up to second order.
Nevertheless, technical details will be omitted for the benefict of the reader. Elaborated calcula can be found in \cite{Steph}.
\medskip

The Lax pair and the non-isospectral condition must be invariant under
($\ref{3.1}$), in order for ($\ref{3.2}$) to be a symmetry vector field. This imposition leads us to an overdetermined
system of equations for the coefficients of the infinitesimal generator $X$.
\bigskip

The classical method for finding Lie symmetries encloses the following steps:

\begin{enumerate}
\item{Calculation of the prolongations of the coefficients of the infinitesimal generator up to first and second order derivatives, for
the case that concerns us.}
\item{Substitution of the transformed fields in the Lax pair and non-isospectral condition.}
\item{Set all the coefficients in $\epsilon$ equal to zero.}
\item{Substitution of the prolongations.}
\item{$\psi_x,\phi_x,\psi_t,\phi_t,\lambda_t$ can be replaced by using the equations (\ref{2.1}), (\ref{2.2}) and (\ref{2.3}).}
\item{A system of equations for the coefficients of the infinitesimal generator arises by setting equal to zero the coefficients accompanying the remaining derivatives.}
\end{enumerate}
\subsection{Classical Lie symmetries for mCH(2+1) spectral problem}

From the above mentioned procedure \cite{Steph}, the symmetries read:
\begin{align}
 &\xi_1=A_1(y),\nonumber \\
&\xi_2=a_2y+b_2,\nonumber \\
 & \xi_3=a_3t+b_3,\nonumber \\
 &\eta_{\lambda}(y,t,\lambda)=\frac{\left(a_2-a_3\right)}{n}\,\lambda,\nonumber \\
 &\eta_u(x,y,t,u)=\frac{\left(a_3-a_2\right)}{2n}\,u,\nonumber \\
 &\eta_\omega^{[j]}(x,y,t,\omega^{[j]})=\delta^{(j,1)}\,\frac{dA_1(y)}{dy}-\frac{(n-j+1)a_2+(j-1)a_3}{n}\,\omega^{[j]},\nonumber \\
 &\eta_v^{[j]}(x,y,t,v^{[j]})=\delta^{(j,n)}\,A_n(y,t)-\frac{(2(n-j)+1)a_2+(2j-1)a_3}{2n}\,v^{[j]},\nonumber \\
 &\eta_{\phi}(x,y,t,\lambda,\psi,\phi)=\gamma(y,t,\lambda)\,\phi,\nonumber \\
 &\eta_{\psi}(x,y,t,\lambda,\psi,\phi)=\gamma(y,t,\lambda)\,\psi, \label{3.3}\end{align}
with $j=1\dots n.$

The functions $A_1(y)$, $A_n(y,t)$ and constants $a_2, a_3,b_2, b_3$ are arbitrary, whereas the function $\gamma(y,t,\lambda)$ satisfies the condition:
\begin{equation}\frac{\partial\, \gamma(y,t,\lambda)}{\partial t}=\lambda^n\,\frac{\partial \,\gamma(y,t,\lambda)}{\partial y}\label{3.4}\end{equation}
and functions $\delta^{(j,1)}$, $\delta^{(j,n)}$ are Kr\"onecker deltas.
The calculation of symmetries is a long and tedious task which relies in the use of a symbolic calculus package, most of the times. In our case, we
have made use of MAPLE14 to handle the resulting and intermediate cumbersome expressions.

\section{Classification of reductions}

The reductions can be achieved by solving the following characteristic system \cite{Steph}:

\begin{equation}
\frac{dx}{\xi_1}=\frac{dy}{\xi_2}=\frac{dt}{\xi_3}=\frac{d\lambda}{\eta_{\lambda}}=\frac{du}{\eta_{u}}=\frac{d\omega^{[j]}}{\eta_{\omega}^{[j]}}
=\frac{dv^{[j]}}{\eta_{v}^{[j]}}=\frac{d\psi}{\eta_{\psi}}=\frac{d\phi}{\eta_{\phi}}\label{4.1}
,\end{equation}
with $j=1\dots n$.

Several reductions emerge for different values of $a_2,a_3,b_2,b_3$.
There are eight non-trivial independent reductions, which are listed below:
\begin{itemize}
\item{Type I: $a_2\neq 0,\quad b_2=0$}
\begin{itemize}
\item{I.1. $a_3\neq 0,\quad  b_3=0$}
\item{I.2. $a_3=0,\quad  b_3\neq 0$}
\item{I.3. $a_3=0,\quad  b_3=0$}
\end{itemize}
\item{Type II: $a_2=0,\quad  b_2 \neq 0$}
\begin{itemize}
\item{II.1. $a_3\neq 0,\quad  b_3=0$}
\item{II.2. $a_3=0,\quad  b_3\neq 0$}
\item{II.3. $a_3=0,\quad  b_3=0$}
\end{itemize}
\item{Type III: $a_2=0,\quad  b_2=0$}
\begin{itemize}
\item{III.1. $a_3\neq 0,\quad  b_3=0$}
\item{III.2. $a_3=0,\quad  b_3\neq 0$}
\end{itemize}
\end{itemize}

\subsection*{\textbf{Reduction I.1. $a_2\neq 0, b_2=0, a_3\neq 0, b_3=0$.}}
Solving the characteristic system (\ref{4.1}), we obtain the following results:

\begin{itemize}
\item Reduced variables:
\begin{equation}z_1=x-\frac{1}{a_2}\,\int{\frac{A_1(y)}{y}\,dy}, \qquad z_2=\frac{t}{y^r},\end{equation}
where $r=\frac{a_3}{a_2}.$
\item{Reduced parameter}:
\begin{equation}\lambda(y,t)=y^{\frac{1-r}{n}}\,\Lambda(z_2),\end{equation} where $\Lambda(z_2)$ satisfies the \textbf{non-isospectral condition}
\begin{equation}\frac{d\Lambda(z_2)}{dz_2}=\left(\frac{1-r}{n}\right)\,\frac{\Lambda(z_2)^{n+1}}{1+rz_2\,\Lambda(z_2)^n}.\end{equation}
\item{Reduced fields}:

\begin{align}
& u(x,y,t)=y^{\frac{r-1}{2n}}\, U(z_1,z_2),\nonumber\\
&   \omega^{[1]}(x,y,t)=\frac{A_1(y)}{a_2\,y}+\frac{\Omega^{[1]}(z_1,z_2)}{y},\nonumber\\
  & \omega^{[j]}(x,y,t)=y^{\frac{(r-1)(1-j)}{n}}\,\left(\frac{\Omega^{[j]}(z_1,z_2)}{y}\right),\quad \quad j=2\dots n,\nonumber\\
&   v^{[j]}(x,y,t)=y^{\frac{(r-1)(1-2j)}{2n}}\,\left(\frac{V^{[j]}(z_1,z_2)}{y}\right),\quad j=1\dots n-1,\nonumber\\
& v^{[n]}(x,y,t)=y^{\frac{(r-1)(1-2n)}{2n}}\,\left(\frac{1}{a_2\,y}\int{\frac{A_n(y,t)\,y^{\frac{2rn-r+1}{2n}}}{y}}+\frac{V^{[n]}(z_1,z_2)}{y}\right),\nonumber\\
&  \phi(x,y,t)=e^{\int{\frac{\Gamma(y,z_2)}{a_2y}\,dy}}\,\Phi(z_1,z_2),\nonumber\\
 & \psi(x,y,t)=e^{\int{\frac{\Gamma(y,z_2)}{a_2y}\,dy}}\,\Psi(z_1,z_2),\nonumber\\
&  p(x,y,t)=\frac{y^{\frac{r-1}{n}}}{y^r}\,\left(\frac{A_1(y)\,\Lambda(z_2)^{n-1}}{a_2}+P(z_1,z_2)\right),\label{4.5}\\
 & q(x,y,t)=\frac{y^{\frac{r-1}{2n}}}{y^r}\,\left({\frac{1}{a_2}\int\frac{A_n(y,t)\,y^{\frac{2rn-r+1}{2n}}}{y}\,dy}+Q(z_1,z_2)\right).\nonumber\end{align}
where: $$P(z_1,z_2)=\sum_{j=1}^n{\Lambda(z_2)^{n-j}\,\Omega^{[j]}(z_1,z_2)},$$
and $$Q(z_1,z_2)=\sum_{j=1}^{n}{\Lambda(z_2)^{n-j}\,V^{[j]}(z_1,z_2)}.$$
The function $\Gamma(y,z_2)$ is obtained through the following identification:
\begin{equation}\Gamma(y,z_2)=\gamma\left(y,t=z_2y^r\right)\end{equation}
that yields:
$$\gamma_t=\frac{\Gamma_{z_2}}{y^r},\quad\quad
\gamma_y=\Gamma_y-r\,\frac{z_2}{y}\,\Gamma_{z_2}.$$
Therefore, according to (\ref{3.4}) $\Gamma(y,z_2)$ satisfies the equation:
\begin{equation} (1+rz_2\Lambda^n)\,\Gamma_{z_2}=y\Lambda^n\,\Gamma_y.\end{equation}
\item{Reduced spectral problem}:
The reduction of (\ref{2.1})-(\ref{2.2}) yields the non-isospectral Lax pair:
\begin{align}
  (1+r\,z_2\,\Lambda^n)\left(\begin{array}{c}
 \Phi\\
 \Psi
\end{array}\right)_{z_2}&=\Lambda P \left(\begin{array}{c}
 \Phi\\
 \Psi
\end{array}\right)_{z_1}\nonumber\\ &+\frac{i\sqrt{\Lambda}}{2}\left[\begin{array}{cc}
 0&Q_{z_1}-Q\\
Q_{z_1}+Q& 0
\end{array}\right]_{z_1}\left(\begin{array}{c}
 \Phi\\
 \Psi
\end{array}\right),\nonumber\\
  \left(\begin{array}{cc}
 \Phi\\
 \Psi
\end{array}\right)_{z_1}&=\frac{1}{2}\left[\begin{array}{cc}
 -1& i\sqrt{\Lambda}\,U\\
 i\sqrt{\Lambda}\,U& 1
\end{array}\right]\left(\begin{array}{cc}
 \Phi\\
 \Psi
\end{array}\right).\label{4.8}
\end{align}

\item{Reduced hierarchy}:
The compatibility condition of (\ref{4.8}) yields the $1+1$ dimensional non-autonomous hierarchy:
\begin{align}
&\left(V^{[n]}_{ z_1z_1}-V^{[n]}\right)_{z_1}- U_{z_2}=0,&\nonumber\\
& V^{[j]}_{ z_1z_1}-V^{[j]}+U\,\Omega^{(j+1)}=0,\quad\quad &j=1\dots n-1,\nonumber\\
&\left(\Omega^{[1]}\,U\right)_{ z_1}+\frac{r-1}{2n}\,U-r\,z_2\,U_{z_2}=0,&\nonumber\\
&\Omega^{[j]}_{z_1}=U\,V^{[j]}_{z_1}.\quad\quad &j=1\dots n.
\label{4.9}\end{align}
\end{itemize}

\subsection*{\textbf{Reduction I.2. $a_2\neq 0, b_2=0, a_3=0, b_3\neq 0$.}}

\begin{itemize}
\item Reduced variables:
\begin{equation}z_1=x-\frac{1}{a_2}\int{\frac{A_1(y)}{y}\,dy}, \qquad z_2=\frac{a_2\,t}{b_3}-\ln(y).\end{equation}
\item{Reduced parameter}:
\begin{equation}\lambda(y,t)=\left(\frac{a_2\,y}{b_3}\right)^{\frac{1}{n}}\Lambda(z_2),\end{equation} with $\Lambda(z_2)$ satisfying the \textbf{non-isospectral condition}
\begin{equation}\frac{d\Lambda(z_2)}{dz_2}=\frac{\Lambda(z_2)^{n+1}}{n\left(1+\Lambda(z_2)^n\right)}.\end{equation}
\item{Reduced fields}:
\begin{align}
 & u(x,y,t)=\left(\frac{a_2\,y}{b_3}\right)^{\frac{-1}{2n}}\,U(z_1,z_2),\nonumber \\
 &   \omega^{[1]}(x,y,t)=\frac{A_1(y)}{a_2\,y}+\frac{\Omega^{[1]}(z_1,z_2)}{y},\nonumber \\
 &   \omega^{[j]}(x,y,t)=\left(\frac{a_2\,y}{b_3}\right)^{\frac{j-1}{n}}\,\frac{\Omega^{[j]}(z_1,z_2)}{y},\quad \quad j=2\dots n,\nonumber \\
 &   v^{[j]}(x,y,t)=\left(\frac{a_2\,y}{b_3}\right)^{\frac{2j-1}{2n}}\,\frac{V^{[j]}(z_1,z_2)}{y},\quad \quad j=1\dots n-1,\nonumber \\
 &   v^{[n]}(x,y,t)=y^{\frac{-1}{2n}}\int{\frac{A_n(y,t)\,y^{\frac{1}{2n}}}{a_2\,y}\,dy}+\left(\frac{a_2 \, y}{b_3}\right)^{\frac{2n-1}{2n}}\frac{V^{[n]}(z_1,z_2)}{y},\nonumber \\
 &   \phi(x,y,t)=e^{\int{\frac{\Gamma(y,z_2)}{a_2y}\,dy}}\,\Phi(z_1,z_2),\nonumber \\
  &  \psi(x,y,t)=e^{\int{\frac{\Gamma(y,z_2)}{a_2y}\,dy}}\,\Psi(z_1,z_2),\nonumber \\
  &  p(x,y,t)=\left(\frac{a_2\, y}{b_3}\right)^{\frac{n-1}{n}}\left(\frac{A_1(y)\,\Lambda(z_2)^{n-1}}{a_2 y}+\frac{P(z_1,z_2)}{y}\right),\nonumber \\
 &   q(x,y,t)=y^{\frac{-1}{2n}}\,\int{\frac{A_n(y,t)\,y^{\frac{1-2n}{2n}}}{a_2}\,dy}+\left(\frac{a_2\,y}{b_3}\right)^{\frac{2n-1}{2n}}\,\frac{Q(z_1,z_2)}{y},\nonumber\\
  &  P(z_1,z_2)=\sum_{j=1}^n{\Lambda(z_2)^{n-j}\,\Omega^{[j]}(z_1,z_2)},\\
  &  Q(z_1,z_2)=\sum_{j=1}^{n}{\Lambda(z_2)^{n-j}\,V^{[j]}(z_1,z_2)},\nonumber\end{align}
where the function $\Gamma(y,z_2)$ is obtained through the following identification:
\begin{equation}\Gamma(y,z_2)=\gamma\left(y,t=\frac{b_3}{a_2}(z_2+\ln(y))\right)\label{4.13}\end{equation}
that yields:
$$\gamma_t=\frac{a_2}{b_3}\,\Gamma_{z_2},\quad\quad
\gamma_y=\Gamma_y-\frac{1}{y}\,\Gamma_{z_2}.$$
Therefore, according to (\ref{3.4}), $\Gamma(y,z_2)$ satisfies the equation:
\begin{equation} (1+\Lambda^n)\,\Gamma_{z_2}=y\Lambda^n\,\Gamma_y.\end{equation}
\item{Reduced spectral problem}:

\begin{align}
  (1+\Lambda^n)\left(\begin{array}{c}
 \Phi\\
 \Psi
\end{array}\right)_{z_2}&=\Lambda P \left(\begin{array}{c}
 \Phi\\
 \Psi
\end{array}\right)_{z_1}\nonumber \\ &+\frac{i\sqrt{\Lambda}}{2}\left[\begin{array}{cc}
 0&Q_{z_1}-Q\\
 Q_{z_1}+Q& 0
\end{array}\right]_{z_1}\left(\begin{array}{c}
 \Phi\\
 \Psi
\end{array}\right),\nonumber\\
  \left(\begin{array}{cc}
 \Phi\\
 \Psi
\end{array}\right)_{z_1}&=\frac{1}{2}\left[\begin{array}{cc}
 -1& i\sqrt{\Lambda}\,U\\
 i\sqrt{\Lambda}\,U& 1
\end{array}\right]\left(\begin{array}{cc}
 \Phi\\
 \Psi
\end{array}\right).
\end{align}
\item{Reduced hierarchy}: This non-isospectral yields the autonomous hierarchy:
\begin{align}
&\left(V^{[n]}_{ z_1z_1}- V^{[n]}\right)_{ z_1}- U_{ z_2} =0,&\nonumber \\
 &V^{[j]}_{ z_1z_1}-V^{[j]}+U\,\Omega^{(j+1)}=0,\quad \quad &j=1\dots n-1,\nonumber \\
 &\left(\Omega^{[1]}\,U\right)_{z_1}-\frac{U}{2n}- U_{ z_2}=0,&\nonumber \\
 &\Omega^{[j]}_{ z_1}=U\,V^{[j]}_{ z_1},\quad \quad &j=1\dots n.
\end{align}
\end{itemize}

\subsection*{\textbf{Reduction I.3. $a_2\neq 0, b_2=0, a_3=0, b_3=0$.}}

\begin{itemize}
\item Reduced variables:
\begin{equation}z_1=x-\frac{1}{a_2}\int{\frac{A_1(y)}{y}\,dy}, \qquad z_2=t.\end{equation}
\item{Reduced parameter}:
\begin{equation}\lambda(y,t)=y^{\frac{1}{n}}\,\Lambda(z_2),\end{equation} where $\Lambda(z_2)$ satisfies the \textbf{non-isospectral condition} \begin{equation}\frac{d\Lambda(z_2)}{dz_2}=\frac{\Lambda(z_2)^{n+1}}{n}.\end{equation}
\item{Reduced fields}:
\begin{align}  &  u(x,y,t)=y^{-\frac{1}{2n}}\,U(z_1,z_2),\nonumber \\
  &  \omega^{[1]}(x,y,t)=\frac{A_1(y)}{a_2y}+\frac{\Omega^{[1]}(z_1,z_2)}{y},\nonumber \\
  &  \omega^{[j]}(x,y,t)=y^{\frac{j-1}{n}}\,\frac{\Omega^{[j]}(z_1,z_2)}{y},\quad \quad j=2\dots n,\nonumber \\
 &   v^{[j]}(x,y,t)=y^{\frac{2j-1}{2n}}\,\frac{V^{[j]}(z_1,z_2)}{y},\quad \quad j=1\dots n-1,\nonumber \\
 &   v^{[n]}(x,y,t)=y^{-\frac{1}{2n}}\,\left(\int{\frac{A_n(y,t)\,y^{\frac{1}{2n}}}{a_2\,y}\,dy}+V^{[n]}(z_1,z_2)\right),\nonumber \\
 &   \phi(x,y,t)=e^{\int{\frac{\Gamma(y,z_2)}{a_2y}\,dy}}\,\Phi(z_1,z_2),\nonumber\\
 &   \psi(x,y,t)=e^{\int{\frac{\Gamma(y,z_2)}{a_2y}\,dy}}\,\Psi(z_1,z_2), \nonumber\\
 &   p(x,y,t)=y^{-\frac{1}{n}}\,\left(\frac{A_1(y)\,\Lambda(z_2)^{n-1}}{a_2}+P(z_1,z_2)\right),\nonumber \\
 &   q(x,y,t)=y^{-\frac{1}{2n}}\,\left(\frac{1}{a_2}\int{\frac{A_n(y,t)\,y^{\frac{1}{2n}}}{y}\,dy}+Q(z_1,z_2)\right),\nonumber\\
 &   P(z_1,z_2)=\sum_{j=1}^n{\Lambda(z_2)^{n-j}\,\Omega^{[j]}(z_1,z_2)},\\
 &   Q(z_1,z_2)=\sum_{j=1}^{n}{\Lambda(z_2)^{n-j}\,V^{[j]}(z_1,z_2)},\nonumber\end{align}
where the function $\Gamma(y,z_2)$ is obtained through the following identification:
\begin{equation}\Gamma(y,z_2)=\gamma\left(y,t=z_2\right)\end{equation}
that yields:
$$\gamma_t=\Gamma_{z_2},\quad\quad
\gamma_y=\Gamma_y.$$
Therefore, according to (\ref{3.4}), $\Gamma(y,z_2)$ satisfies the equation:
\begin{equation} \Gamma_{z_2}=y\Lambda^n\,\Gamma_y.\end{equation}
\item{Reduced spectral problem}:
\begin{align}
  \left(\begin{array}{c}
 \Phi\\
 \Psi
\end{array}\right)_{z_2}&=\Lambda P \left(\begin{array}{c}
 \Phi\\
 \Psi
\end{array}\right)_{z_1}\nonumber \\&+\frac{i\sqrt{\Lambda}}{2}\left[\begin{array}{cc}
 0&Q_{z_1}-Q\\
 Q_{z_1}+Q& 0
\end{array}\right]_{z_1}\left(\begin{array}{c}
 \Phi\\
 \Psi
\end{array}\right),\nonumber\\
  \left(\begin{array}{cc}
 \Phi\\
 \Psi
\end{array}\right)_{z_1}&=\frac{1}{2}\left[\begin{array}{cc}
 -1& i\sqrt{\Lambda}\,U\\
 i\sqrt{\Lambda}\,U& 1
\end{array}\right]\left(\begin{array}{cc}
 \Phi\\
 \Psi
\end{array}\right).
\end{align}
\item{Reduced hierarchy}:
\begin{align}
&V^{[n]}_{z_1z_1z_1}- V^{[n]}_{ z_1}-U_{z_2}=0,\nonumber &\\
& V^{[j]}_{z_1z_1}-V^{[j]}+U\,\Omega^{(j+1)}=0,\quad \quad & j=1\dots n-1,\nonumber\\
&\left(\Omega^{[1]}\,U\right)_{z_1}-\frac{U}{2n}=0, &\nonumber\\
&\Omega^{[j]}_{z_1}=UV^{[j]}_{z_1},\quad \quad & j=1\dots n.\end{align}
\end{itemize}

\subsection*{\textbf{Reduction II.1. $a_2=0, b_2 \neq 0, a_3\neq 0, b_3=0.$}}
\begin{itemize}
\item Reduced variables:
\begin{equation}z_1=x-\frac{1}{b_2}\int A_1(y)\,dy, \qquad z_2=\frac{a_3\, t}{b_2}\,e^{\frac{-a_3y}{b_2}}.\end{equation}
\item{Reduced parameter}:
\begin{equation}\lambda(y,t)=e^{\frac{-a_3y}{n\,b_2}}\,\Lambda(z_2),\end{equation} where $\Lambda(z_2)$ satisfies the \textbf{non-isospectral condition} : \begin{equation}\frac{d\Lambda(z_2)}{dz_2}=-\frac{\Lambda(z_2)^{n+1}}{n\left(1+z_2\Lambda(z_2)^n\right)}.\end{equation}
\item{Reduced fields}:
\begin{align}  &   u(x,y,t)=e^{\frac{a_3y}{2n\,b_2}}\,U(z_1,z_2),\nonumber \\
 &   \omega^{[1]}(x,y,t)=\frac{A_1(y)}{b_2}+\frac{a_3}{b_2}\,\Omega^{[1]}(z_1,z_2),\nonumber \\
  &  \omega^{[j]}(x,y,t)=\frac{a_3}{b_2}\,e^{\frac{-a_3(j-1)y}{n\,b_2}}\,\Omega^{[j]}(z_1,z_2),\quad \quad j=2\dots n,\nonumber \\
  &  v^{[j]}(x,y,t)=\frac{a_3}{b_2}\,e^{\frac{-a_3(2j-1)y}{2n\,b_2}}\,V^{[j]}(z_1,z_2),\quad \quad j=1\dots n-1,\nonumber \\
 &   v^{[n]}(x,y,t)=e^{\frac{-a_3(2n-1)y}{2n\,b_2}}\left(\int   \frac{A_n(y,t)\,e^{\frac{a_3(2n-1)y}{2n\,b_2}}}{b_2}\,dy+\frac{a_3}{b_2}\,V^{[n]}(z_1,z_2)\right),\nonumber \\
 &   \phi(x,y,t)=e^{\int\frac{\Gamma(y,z_2)}{b_2}\,dy}\,\Phi(z_1,z_2),\nonumber \\
 &   \psi(x,y,t)=e^{\int\frac{\Gamma(y,z_2)}{b_2}\,dy}\,\Psi(z_1,z_2),\nonumber \\
 &   p(x,y,t)=e^{\frac{-a_3(n-1)y}{n\,b_2}}\,\left(\frac{A_1(y)\,\Lambda(z_2)^{n-1}}{b_2}+\frac{a_3}{b_2}\,P(z_1,z_2)\right),\nonumber \\
 &   q(x,y,t)=e^{\frac{-a_3(2n-1)y}{2n\,b_2}}\left(\int \frac{A_n(y,t)\,e^{\frac{a_3(2n-1)y}{2n\,b_2}}}{b_2}\,dy+\frac{a_3}{b_2}\,Q(z_1,z_2)\right),\nonumber\\
 &   P(z_1,z_2)=\sum_{j=1}^n{\Lambda(z_2)^{n-j}\,\Omega^{[j]}(z_1,z_2)},\\
 &   Q(z_1,z_2)=\sum_{j=1}^{n}{\Lambda(z_2)^{n-j}\,V^{[j]}(z_1,z_2)},\nonumber\nonumber\end{align}
where the function $\Gamma(y,z_2)$ is obtained through the following identification:
\begin{equation}\Gamma(y,z_2)=\gamma\left(y,t=\frac{b_2}{a_3}\,e^{\frac{a_3}{b_2}}\right)\end{equation}
that yields:
$$\gamma_t=\frac{a_3}{b_2}\,e^{\frac{-a_3y}{b_2}}\,\Gamma_{z_2},\quad\quad
\gamma_y=\Gamma_y-\frac{a_3}{b_2}z_2\,\Gamma_{z_2}.$$
Therefore, according to (\ref{3.4}), $\Gamma(y,z_2)$ satisfies the equation:
\begin{equation} a_3(1+z_2\Lambda^n)\Gamma_{z_2}=b_2\Lambda^n\,\Gamma_y.\end{equation}
\item{Reduced spectral problem}:
\begin{align}
  (1+z_2\Lambda^n)\left(\begin{array}{c}
 \Phi\\
 \Psi
\end{array}\right)_{z_2}&=\Lambda P \left(\begin{array}{c}
 \Phi\\
 \Psi
\end{array}\right)_{z_1}\nonumber \\ &+\frac{i\sqrt{\Lambda}}{2}\left[\begin{array}{cc}
 0&Q_{z_1}-Q\\
 Q_{z_1}+Q& 0
\end{array}\right]_{z_1}\left(\begin{array}{c}
 \Phi\\
 \Psi
\end{array}\right),\nonumber\\
  \left(\begin{array}{cc}
 \Phi\\
 \Psi
\end{array}\right)_{z_1}&=\frac{1}{2}\left[\begin{array}{cc}
 -1& i\sqrt{\Lambda}\,U\\
 i\sqrt{\Lambda}\,U& 1
\end{array}\right]\left(\begin{array}{cc}
 \Phi\\
 \Psi
\end{array}\right).
\end{align}
\item{Reduced hierarchy}:

\begin{align}
& V^{[n]}_{z_1z_1z_1}- V^{[n]}_{ z_1}-U_{z_2}=0, &\nonumber\\
& V^{[j]}_{z_1z_1}-V^{[j]}+U\,\Omega^{(j+1)}=0,\quad \quad & j=1\dots n-1,\nonumber\\
& \left(\Omega^{[1]}\,U\right)_{z_1}+\frac{U}{2n}-z_2U_{z_2}=0, & \nonumber\\
& \Omega^{[j]}_{z_1}=UV^{[j]}_{z_1},\quad \quad  & j=1\dots n.\end{align}
\end{itemize}

\subsection*{\textbf{Reduction II.2.} $a_2=0, b_2 \neq 0, a_3=0, b_3\neq 0$.}

\begin{itemize}
\item Reduced variables:
$$z_1=x-\int{\frac{A_1(y)}{b_2}\,dy}, \qquad z_2=y-\frac{b_2}{b_3}t.$$
\item{Reduced parameter}:
$$\lambda(y,t)=\left(\frac{b_2}{b_3}\right)^{\frac{1}{n}}\Lambda(z_2),$$ where $\Lambda(z_2)$ satisfies the \textbf{isospectral condition}:
$$\frac{d\Lambda(z_2)}{dz_2}=0.$$
\item{Reduced fields}:
\begin{align}   &  u(x,y,t)=\left(\frac{b_3}{b_2}\right)^{\frac{1}{2n}}\,U(z_1,z_2),\nonumber \\
  &  \omega^{[1]}(x,y,t)=\frac{A_1(y)}{b_2}+\Omega^{[1]}(z_1,z_2),\nonumber \\
 &   \omega^{[j]}(x,y,t)=\left(\frac{b_2}{b_3}\right)^{\frac{j-1}{n}}\,\Omega^{[j]}(z_1,z_2),\quad \quad j=2\dots n,\nonumber \\
 &   v^{[j]}(x,y,t)=\left(\frac{b_2}{b_3}\right)^{\frac{2j-1}{2n}}\,V^{[j]}(z_1,z_2),\quad \quad j=1\dots n-1,\nonumber \\
 &   v^{[n]}(x,y,t)=\frac{1}{b_2}\int A_n(y,t)\,dy+\left(\frac{b_2}{b_3}\right)^{\frac{2n-1}{2n}}\,V^{[n]}(z_1,z_2),\nonumber \\
 &   \phi(x,y,t)=e^{\int\frac{\Gamma(y,z_2)}{b_2}\,dy}\,\Phi(z_1,z_2),\nonumber \\
 &   \psi(x,y,t)=e^{\int\frac{\Gamma(y,z_2)}{b_2}\,dy}\,\Psi(z_1,z_2), \nonumber\\
 &   p(x,y,t)=\left(\frac{b_2}{b_3}\right)^{\frac{n-1}{n}}\,\left(\frac{A_1(y)\Lambda(z_2)^{n-1}}{b_2}+P(z_1,z_2)\right),\nonumber \\
  &  q(x,y,t)=\frac{1}{b_2}\int A_n(y,t)\,dy+\left(\frac{b_2}{b_3}\right)^{\frac{2n-1}{2n}}Q(z_1,z_2) .\nonumber\\
 &   P(z_1,z_2)=\sum_{j=1}^n{\Lambda(z_2)^{n-j}\,\Omega^{[j]}(z_1,z_2)},\\
 &   Q(z_1,z_2)=\sum_{j=1}^{n}{\Lambda(z_2)^{n-j}\,V^{[j]}(z_1,z_2)},\nonumber\end{align}
where the function $\Gamma(y,z_2)$ is obtained through the following identification:
\begin{equation}\Gamma(y,z_2)=\gamma\left(y,t=\frac{b_3}{b_2}(y-z_2)\right)\end{equation}
that yields:
$$\gamma_t=-\frac{b_2}{b_3}\,\Gamma_{z_2},\quad\quad\gamma_y=\Gamma_y+\Gamma_{z_2}.$$
Therefore, according to (\ref{3.4}), $\Gamma(y,z_2)$ satisfies the equation:
\begin{equation} (1+\Lambda^n)\Gamma_{z_2}=-\Lambda^n\,\Gamma_y.\end{equation}
\item{Reduced spectral problem}:
\begin{align}
  (1+\Lambda^n)\left(\begin{array}{c}
 \Phi\\
 \Psi
\end{array}\right)_{z_2}&=\Lambda P \left(\begin{array}{c}
 \Phi\\
 \Psi
\end{array}\right)_{z_1}\nonumber \\ &+\frac{i\sqrt{\Lambda}}{2}\left[\begin{array}{cc}
 0&Q_{z_1}-Q\\
 Q_{z_1}+Q& 0
\end{array}\right]_{z_1}\left(\begin{array}{c}
 \Phi\\
 \Psi
\end{array}\right),\nonumber\\
  \left(\begin{array}{cc}
 \Phi\\
 \Psi
\end{array}\right)_{z_1}&=\frac{1}{2}\left[\begin{array}{cc}
 -1& i\sqrt{\Lambda}\,U\\
 i\sqrt{\Lambda}\,U& 1
\end{array}\right]\left(\begin{array}{cc}
 \Phi\\
 \Psi
\end{array}\right).
\end{align}
\item{Reduced hierarchy}:
\begin{align}
& V^{[n]}_{z_1z_1z_1}- V^{[n]}_{ z_1}+U_{z_2}=0, &\nonumber\\
& V^{[j]}_{z_1z_1}-V^{[j]}+U\,\Omega^{(j+1)}=0,\quad \quad & j=1\dots n-1,\nonumber\\
& \left(\Omega^{[1]}\,U\right)_{z_1}+U_{z_2}=0,& \nonumber\\
&  \Omega^{[j]}_{z_1}=UV^{[j]}_{z_1},\quad \quad & j=1\dots n.\end{align}
\end{itemize}

\subsection*{\textbf{Reduction II.3. $a_2=0, b_2 \neq 0, a_3=0, b_3=0$.}}

\begin{itemize}
\item Reduced variables:
$$z_1=x-\int{\frac{A_1(y)}{b_2}\,dy}, \qquad z_2=t.$$
\item{Reduced parameter}:
$$\lambda(y,t)=\Lambda(z_2),$$ where $\Lambda(z_2)$ satisfies the \textbf{isospectral condition}:
$$\frac{d\Lambda(z_2)}{dz_2}=0.$$
\item{Reduced fields}:
\begin{align}   &  u(x,y,t)=U(z_1,z_2),\nonumber \\
 &   \omega^{[1]}(x,y,t)=\frac{A_1(y)}{b_2}+\Omega^{[1]}(z_1,z_2),\nonumber \\
 &   \omega^{[j]}(x,y,t)=\Omega^{[j]}(z_1,z_2),\quad \quad j=2\dots n,\nonumber \\
 &   v^{[j]}(x,y,t)=V^{[j]}(z_1,z_2),\quad \quad j=1\dots n-1,\nonumber \\
 &   v^{[n]}(x,y,t)=\frac{1}{b_2}\int A_n(y,t)\,dy+V^{[n]}(z_1,z_2),\nonumber \\
 &   \phi(x,y,t)=e^{\int\frac{\Gamma(y,z_2)}{b_2}\,dy}\,\Phi(z_1,z_2),\nonumber \\
 &   \psi(x,y,t)=e^{\int\frac{\Gamma(y,z_2)}{b_2}\,dy}\,\Psi(z_1,z_2),\nonumber \\
 &   p(x,y,t)=\frac{A_1(y)\Lambda(z_2)^{n-1}}{b_2}+P(z_1,z_2),\nonumber \\
 &   q(x,y,t)=\frac{1}{b_2}\int A_n(y,t)\,dy+Q(z_1,z_2),\nonumber\\
  &  P(z_1,z_2)=\sum_{j=1}^n{\Lambda(z_2)^{n-j}\,\Omega^{[j]}(z_1,z_2)},\\
  &  Q(z_1,z_2)=\sum_{j=1}^{n}{\Lambda(z_2)^{n-j}\,V^{[j]}(z_1,z_2)},\nonumber\end{align}
where the function $\Gamma(y,z_2)$ is obtained through the following identification:
\begin{equation}\Gamma(y,z_2)=\gamma\left(y,t=z_2\right)\end{equation}
that yields:
$$\gamma_t=\Gamma_{z_2},\quad\quad
\gamma_y=\Gamma_y.$$
Therefore, according to (\ref{3.4}), $\Gamma(y,z_2)$ satisfies the equation:
\begin{equation}\Gamma_{z_2}=\Lambda^n\,\Gamma_y.\end{equation}
\item{Reduced spectral problem}:
\begin{align}
  \left(\begin{array}{c}
 \Phi\\
 \Psi
\end{array}\right)_{z_2}&=\Lambda P \left(\begin{array}{c}
 \Phi\\
 \Psi
\end{array}\right)_{z_1}\nonumber \\ &+\frac{i\sqrt{\Lambda}}{2}\left[\begin{array}{cc}
 0&Q_{z_1}-Q\\
 Q_{z_1}+Q& 0
\end{array}\right]_{z_1}\left(\begin{array}{c}
 \Phi\\
 \Psi
\end{array}\right),\nonumber\\
  \left(\begin{array}{cc}
 \Phi\\
 \Psi
\end{array}\right)_{z_1}&=\frac{1}{2}\left[\begin{array}{cc}
 -1& i\sqrt{\Lambda}\,U\\
 i\sqrt{\Lambda}\,U& 1
\end{array}\right]\left(\begin{array}{cc}
 \Phi\\
 \Psi
\end{array}\right).
\end{align}
\item{Reduced hierarchy}:
\begin{align}
& V^{[n]}_{z_1z_1z_1}- V^{[n]}_{ z_1}-U_{z_2}=0, &\nonumber\\
& V^{[j]}_{z_1z_1}-V^{[j]}+U\,\Omega^{(j+1)}=0,\quad \quad & j=1\dots n-1,\nonumber\\
& \left(\Omega^{[1]}\,U\right)_{z_1}=0, & \nonumber\\
& \Omega^{[j]}_{z_1}=UV^{[j]}_{z_1},\quad \quad  & j=1\dots n.\end{align}
\end{itemize}
\subsection*{\textbf{Reduction III.1.} $a_2=0, b_2=0, a_3\neq 0, b_3=0$.}
\begin{itemize}
\item Reduced variables:
$$z_1=x-\frac{A_1(y)}{a_3}\,\ln t, \qquad z_2=y.$$
\item{Reduced parameter}:
$$\lambda(y,t)=t^{\frac{-1}{n}}\,\Lambda(z_2),$$ where $\Lambda(z_2)$ satisfies the \textbf{nonisospectral condition}:
$$n\frac{d\Lambda(z_2)}{dz_2}+\Lambda(z_2)^{1-n}=0.$$
\item{Reduced fields}:
\begin{align}  &   u(x,y,t)=t^{\frac{1}{2n}}\,U(z_1,z_2),\nonumber \\
  &  \omega^{[1]}(x,y,t)=\frac{\ln(t)}{a_3}\,\frac{d\,A_1(y)}{dy}+\Omega^{[1]}(z_1,z_2),\nonumber \\
  &  \omega^{[j]}(x,y,t)=t^{\frac{1-j}{n}}\,\Omega^{[j]}(z_1,z_2),\quad \quad j=2\dots n,\nonumber \\
 &   v^{[j]}(x,y,t)=t^{\frac{1-2j}{2n}}\,V^{[j]}(z_1,z_2),\quad \quad j=1\dots n-1,\nonumber \\
 &   v^{[n]}(x,y,t)=t^{\frac{1-2n}{2n}}\,\left(\frac{1}{a_3}\int t^{\frac{-1}{2n}}\,A_n(y,t)\,dt+V^{[n]}(z_1,z_2)\right),\nonumber \\
 &   \phi(x,y,t)=e^{\int\frac{\Gamma(t,z_2)}{a_3t}\,dt}\,\Phi(z_1,z_2),\nonumber \\
 &   \psi(x,y,t)=e^{\int\frac{\Gamma(t,z_2)}{a_3t}\,dt}\,\Psi(z_1,z_2), \nonumber\\
 &   p(x,y,t)=t^{\frac{1-n}{n}}\,\left(\frac{\ln(t)\,\Lambda(z_2)^{n-1}}{a_3}\,\frac{d\,A_1(y)}{dy}+P(z_1,z_2)\right),\nonumber \\
 &   q(x,y,t)=t^{\frac{1-2n}{2n}}\,\left(\frac{1}{a_3}\int t^{\frac{-1}{2n}}\,A_n(y,t)\,dt+Q(z_1,z_2)\right),\nonumber\\
 &   P(z_1,z_2)=\sum_{j=1}^n{\Lambda(z_2)^{n-j}\,\Omega^{[j]}(z_1,z_2)},\\
 &   Q(z_1,z_2)=\sum_{j=1}^{n}{\Lambda(z_2)^{n-j}\,V^{[j]}(z_1,z_2)},\nonumber\end{align}
and the function $\Gamma(y,z_2)$ is obtained through the following identification:
\begin{equation}\Gamma(t,z_2)=\gamma\left(y=z_2,t\right)\end{equation}
that yields:
$$\gamma_t=\Gamma_{t},\quad\quad
\gamma_y=\Gamma_{z_2}.$$
Therefore, according to (\ref{3.4}), $\Gamma(y,z_2)$ satisfies the equation
\begin{equation}\Lambda^n\,\Gamma_{z_2}=t\,\Gamma_t.\end{equation}
\item{Reduced spectral problem}:

\begin{align}
-\Lambda^n\left(\begin{array}{c}
 \Phi\\
 \Psi
\end{array}\right)_{z_2}&=\left(\Lambda P +\frac{\hat A_1}{a_3}\right)\left(\begin{array}{c}
 \Phi\\
 \Psi
\end{array}\right)_{z_1}\nonumber \\ &+\frac{i\sqrt{\Lambda}}{2}\left[\begin{array}{cc}
 0&Q_{z_1}-Q\\
 Q_{z_1}+Q& 0
\end{array}\right]_{z_1}\left(\begin{array}{c}
 \Phi\\
 \Psi
\end{array}\right),\nonumber\\
\left(\begin{array}{cc}
 \Phi\\
 \Psi
\end{array}\right)_{z_1}&=\frac{1}{2}\left[\begin{array}{cc}
 -1& i\sqrt{\Lambda}\,U\\
 i\sqrt{\Lambda}\,U& 1
\end{array}\right]\left(\begin{array}{cc}
 \Phi\\
 \Psi
\end{array}\right).
\end{align}
and $\hat A_1=\hat A_1(z_2)=A_1(y=y(z_2)).$
\item{Reduced hierarchy}:
\begin{align}
& V^{[n]}_{z_1z_1z_1}- V^{[n]}_{ z_1}-\frac{U}{2n}+\frac{\hat A_1(z_2)}{a_3}U_{z_1}=0, &\nonumber\\
&  V^{[j]}_{z_1z_1}-V^{[j]}+U\,\Omega^{(j+1)}=0,\quad \quad & j=1\dots n-1,\nonumber\\
& \left(\Omega^{[1]}\,U\right)_{z_1}+U_{z_2}=0, &\nonumber\\
&  \Omega^{[j]}_{z_1}=UV^{[j]}_{z_1},\quad \quad & j=1\dots n.\end{align}
\end{itemize}
\subsection*{\textbf{Reduction III.2.} $a_2=0, b_2=0, a_3=0, b_3\neq 0$.}
\begin{itemize}
\item Reduced variables:
$$z_1=x-\frac{A_1(y)\,t}{b_3}, \qquad z_2=\frac{1}{b_3}\,\int A_1(y)\,dy.$$
\item{Reduced parameter}:
$$\lambda(y,t)=\Lambda(z_2),$$ where $\Lambda(z_2)$ satisfies the \textbf{isospectral condition}:
$$\frac{d\Lambda(z_2)}{dz_2}=0.$$
\item{Reduced fields}:
\begin{align}  &   u(x,y,t)=U(z_1,z_2),\nonumber \\
  &  \omega^{[1]}(x,y,t)=\frac{t}{b_3}\frac{dA_1(y)}{dy}+\frac{A_1(y)}{b_3}\,\Omega^{[1]}(z_1,z_2),\nonumber \\
 &   \omega^{[j]}(x,y,t)=\frac{A_1(y)}{b_3}\,\Omega^{[j]}(z_1,z_2),\quad \quad j=2\dots n,\nonumber \\
  &  v^{[j]}(x,y,t)=\frac{A_1(y)}{b_3}\,V^{[j]}(z_1,z_2),\quad \quad j=1\dots n-1,\nonumber \\
  &  v^{[n]}(x,y,t)=\frac{1}{b_3}\int A_n(y,t)\,dt+\frac{A_1(y)}{b_3}\,V^{[n]}(z_1,z_2),\nonumber \\
  &  \phi(x,y,t)=e^{\int\frac{\Gamma(t,z_2)}{b_3}\,dt}\,\Phi(z_1,z_2),\nonumber \\
 &   \psi(x,y,t)=e^{\int\frac{\Gamma(t,z_2)}{b_3}\,dt}\,\Psi(z_1,z_2), \nonumber\\
 &   p(x,y,t)=\frac{t}{b_3}\,\frac{dA_1(y)}{dy}\,\Lambda^{n-1}+\frac{A_1(y)}{b_3}\,P(z_1,z_2),\nonumber \\
  &  q(x,y,t)=\frac{1}{b_3}\int A_n(y,t)\,dt+\frac{A_1(y)}{b_3}\,Q(z_1,z_2).\nonumber\\
 &   P(z_1,z_2)=\sum_{j=1}^n{\Lambda(z_2)^{n-j}\,\Omega^{[j]}(z_1,z_2)},\\
  &  Q(z_1,z_2)=\sum_{j=1}^{n}{\Lambda(z_2)^{n-j}\,V^{[j]}(z_1,z_2)},\nonumber\end{align}
where the function $\Gamma(y,z_2)$ is obtained through the following identification:
\begin{equation}\Gamma(t,z_2)=\gamma\left(y=y(z_2),t\right)\end{equation}
that yields:
$$\gamma_t=\Gamma_{t},\quad\quad\gamma_y=\frac{A_1}{a_3}\Gamma_{z_2}.$$
Therefore, according to (\ref{3.4}), $\Gamma(y,z_2)$ satisfies the equation:
\begin{equation}\Lambda^n\,\Gamma_{z_2}=\frac{b_3}{\hat A_1(z_2)}\Gamma_t,\quad \hat A_1(z_2)=A_1(y=y(z_2)).\end{equation}
\item{Reduced spectral problem}:
\begin{align}
  -\Lambda^n\left(\begin{array}{c}
 \Phi\\
 \Psi
\end{array}\right)_{z_2}&=\left(\Lambda P +1\right)\left(\begin{array}{c}
 \Phi\\
 \Psi
\end{array}\right)_{z_1}\nonumber\\&+\frac{i\sqrt{\Lambda}}{2}\left[\begin{array}{cc}
 0&Q_{z_1}-Q\\
 Q_{z_1}+Q& 0
\end{array}\right]_{z_1}\left(\begin{array}{c}
 \Phi\\
 \Psi
\end{array}\right),\nonumber\\
 \left(\begin{array}{cc}
 \Phi\\
 \Psi
\end{array}\right)_{z_1}&=\frac{1}{2}\left[\begin{array}{cc}
 -1& i\sqrt{\Lambda}\,U\\
 i\sqrt{\Lambda}\,U& 1
\end{array}\right]\left(\begin{array}{cc}
 \Phi\\
 \Psi
\end{array}\right).
\end{align}
\item{Reduced hierarchy}:
\begin{align}
& V^{[n]}_{z_1z_1z_1}- V^{[n]}_{ z_1}+U_{z_1}=0,& \nonumber\\
&  V^{[j]}_{z_1z_1}-V^{[j]}+U\,\Omega^{(j+1)}=0,\quad \quad  & j=1\dots n-1,\nonumber\\
& \left(\Omega^{[1]}\,U\right)_{z_1}+U_{z_2}=0, &\nonumber\\
&  \Omega^{[j]}_{z_1}=UV^{[j]}_{z_1},\quad \quad & j=1\dots n.\end{align}
\end{itemize}

\section{Conclusions}

We have searched for classical Lie point symmetries of the spectral problem associated with mCH(2+1). These symmetries depend on four arbitrary constants $a_2, b_2, a_3, b_3$ and two arbitrary functions $A_1(y)$ and $A_n(y,t)$,  whereas an additional function $\gamma(y,t,\lambda)$ obeys condition (\ref{2.3}).
 Each similarity reduction leads to a $1+1$ dimensional  spectral problem, whose compatibility condition yields a reduced hierarchy in $1+1$ dimensions.

 All the possible non-trivial reductions of such spectral problem have been summarized in eight different cases. We have explicitly derived these reductions and we have identified eight new integrable hierarchies in 1+1 dimensions. We consider of particular interest observing \textbf{how the spectral parameter and the eigenfunction reduce under the symmetry}. For six of the cases, the reduced spectral problem is non-isospectral. It is important to remark that non-isospectral problems in $1+1$ dimensions are not frequent. These results confirm the group interpretation of the spectral parameter proposed by other authors. \cite{levi1}, \cite{levi2}, \cite{cies1}, \cite{cies2}, \cite{cies3}, \cite{cies4}.

\section*{Acknowledgements}
This research has been supported in part by the DGICYT under project FIS2009-07880. We also thank the referee for suggesting important references.

\end{document}